\documentstyle[prl,aps]{revtex}

\begin{document}
\title{Chaotic Behavior in Shell Models and Shell Maps
} 
\author{Julien Kockelkoren\footnote{Electronic Address: julienk@sci.kun.nl}$^1$,
Fridolin Okkels\footnote{Electronic Address: okkels@nbi.dk}
and Mogens H. Jensen\footnote{Electronic Address: mhjensen@nbi.dk}}
\address{Niels Bohr Institute 
and Center for Chaos and Turbulence Studies, 
Blegdamsvej 17, DK-2100 Copenhagen {\O}, Denmark }
\date{January 19, 1997}
\maketitle
\begin{abstract}
We study the chaotic behavior of the ``GOY'' shell model by measuring
the variation of the maximal Lyapunov exponent with the parameter 
$\epsilon$ which determines the nature of the second invariant
(the generalized ``helicity'' invariant). After a Hopf bifurcation,
we observe a critical point at $\epsilon_c \sim 0.38704$ above which
the maximal Lyapunov exponent grows nearly linearly. 
For high values of $\epsilon$ the evolution becomes regular again
which can be explained by a simple analytic argument. 
A model with few shells shows two transitions.  
To simplify the model
substantially we introduce a shell map which exhibits similar properties
as the``GOY'' model.
\end{abstract}
\pacs{PACS numbers: }

Shell models of turbulence have been studied intensively by Kadanoff
\cite{kada1,kada2,Lohse} and many others (for a review see \cite{mogens}). 
It appears that a lot of properties of highly turbulent flows
are nicely captured by those models and 
numerical computations of shell models are much more tractable than
direct simulations of the Navier-Stokes equations. Of particular focus
have been studies of intermittency effects where laminar quiescent
periods are interrupted by strong intermittency bursts of high
energy dissipation. Most shell models follow completely
deterministic dynamics but exhibits nevertheless strong intermittent
behavior and this is believed to be caused by the internal chaotic
dynamics. For the ``GOY'' shell model \cite{Gledzer,OY}, 
the strength of this chaos is
known to be related to the nature of the second invariant,
i.e. the ``helicity'' invariant \cite{kada1,bif}. It is the purpose
of this paper to explore the nature of the chaotic behavior as 
the properties of the second invariant are changed. We do this by estimating
the variation of the maximal Lyapunov exponent with an external parameter
$\epsilon$ and find a transition in this exponent where it jumps up from zero 
to a small but finite value after which it grows linearly. Previous
work by Biferale et al \cite{bif} and Kadanoff et al \cite{kada2} have
concentrated on the instability of the ``Kolmogorov fixed point'' as
the nature of the invariant is changed, measured by a
parameter $\epsilon$ to be introduced below. The first group
observed a Hopf bifurcation 
of the fixed point at $\epsilon = 0.3843$ turning into a torus 
at $\epsilon = 0.3953$ and finally a strange attractor at $\epsilon = 0.398$. 
This was refined by the other group who also
studied the detailed eigenvalue spectrum for the unstable modes \cite{kada2}.
These studies however only addressed the instability of the
stationary state and not the transition to chaotic evolution,
which is our purpose here.

Shell models are formed by various truncation techniques 
of the Navier-Stokes equations \cite{mogens}. The
most well-studied model is the ``GOY'' model of 
Gledzer-Ohkitani-Yamada 
\cite{kada1,kada2,Lohse,mogens,Gledzer,OY,bif,JPV,benzi,pisarenko}. 
This model yields corrections to the Kolmogorov theory \cite{JPV} in good 
agreement with experiments \cite{Anselmet,vdWater}. 
For the ``GOY'' shell model, wave-number space is divided into
$N$ separated shells each characterized by a wave-number
$k_n=r^n \, k_0$ ($r=2)$,  with $n=1,\cdots N$. 
Each shell is assigned a complex amplitude $u_n$ describing
the typical velocity gradient over a scale $\ell_n = 1/k_n$.
By assuming interactions among 
nearest and next nearest neighbour shells and phase space
volume conservation one arrives at the following 
evolution equations \cite{OY}
\begin{equation}
\label{un}
(\frac{d}{ dt}+\nu k_n^2 ) \ u_n \ = 
 i \,k_n (a_n \,   u^*_{n+1} u^*_{n+2} \, + \, \frac{b_n}{2} 
u^*_{n-1} u^*_{n+1} \, + \,
 \frac{c_n }{4} \,   u^*_{n-1} u^*_{n-2})  \ + \ f \delta_{n,4},
\end{equation}
with boundary conditions $b_1=b_N=c_1=c_2=a_{N-1}=a_N=0$.
$f$  is an external, constant forcing, here on the forth mode.

The  coefficients of the non-linear terms  must follow the relation
$a_n+b_{n+1}+c_{n+2}=0$ in order  to satisfy the conservation  of  energy,
$E = \sum_n |u_n|^2$, when $f=\nu = 0$. 
The constraints still leave a free parameter $\epsilon$ so that 
 one can set
$ a_n=1,\ b_{n+1}=-\epsilon,\ c_{n+2}=-(1-\epsilon)$ \cite{bif}. As observed
by Kadanoff, one obtains the canonical value $\epsilon= 1/2$, if 
helicity conservation is also demanded \cite{kada1}. 
The set (\ref{un}) of $N$ coupled
ordinary differential equations can be numerically integrated by
standard techniques.

To compute the maximal Lyapunov exponent in the GOY model, 
we introduce the notation 
${\bf U} \equiv(Re(u_1),Im(u_1),\cdots,Re(u_N), Im(u_N) \, )$
and $F_i = d U_i/ dt$ and
consider the linear variational equations
\begin{equation}
\label{eq2}
\frac{d z_i}{ dt} \ = \sum_{j=1}^{2N}
    A_{ij} \cdot z_j \qquad \ i \ = \ 1,...,2N
\end{equation}
for the time evolution of an infinitesimal increment ${\bf z}=
 {\bf{\delta U}}$,
 where
\begin{equation}
\label{eq3}
A_{nj} \equiv \partial F_n  /  \partial U_j
\end{equation}
 is the Jacobian matrix of Eqs. (\ref{un}). The solution for the
  tangent vector ${\bf z}$
 can thus be  formally written  as ${\bf z}(t_2)={\bf M}(t_1,t_2) 
\cdot {\bf z}(t_1)$,
 with ${\bf M}= \exp \int_{t_1}^{t_2} \ {\bf A}(\tau) d\tau$.
A generic tangent vector ${\bf z}(t)$
is projected by the evolution along the eigenvector ${\bf e}^{(1)}$, belonging
to the maximum Lyapunov exponent, i.e.
 ${\bf z}(t) = |{\bf z}(0)| \ {\bf e}^{(1)} \exp(\lambda_1 \ t)$ leading to
\begin{equation}
\label{eq4}
\lambda_1 ~=~ \lim_{t \to \infty} \frac{1}{t} {\rm ln}
\frac{| {\bf z}(t) |}{| {\bf z} (0)|}
\end{equation}
where ${\bf z} (0)$ is the initial tangent vector.

Practically, Eqs.\ref{un}, \ref{eq2} are integrated simultaneously over a certain
time $\delta t$, starting with a normalized tangent vector in a random
direction, $\hat{\bf z}(0)$. The increment over time $\delta t$ in the
length of the tangent vector is then
$\delta z_1 = | {\bf z}(\delta t) | / |\hat{\bf z} (0)|$. 
Next, the tangent vector is 
normalized $ \hat{\bf z} (\delta t) = {\bf z} (\delta t) 
/ | {\bf z}(\delta t) |$ and this
vector is used as a seed for a new integration over the time $\delta t$
i.e. propagated forward to $t = 2 \delta t$. Generalizing this
argument we obtain the i'th increment
$\delta z_i = | {\bf z} (i \delta t) | / | 
\hat{\bf z} ((i-1) \delta t) |$
and the maximal Lyapunov exponent is given by
(where we now set $\lambda = \lambda_1$):
\begin{equation}
\label{eq5}
\lambda = \lim_{M \to \infty} 
\frac{1}{M}   \sum_{i=1}^M \frac{{\rm ln}(\delta z_i)}{\delta t}
\end{equation}

We have estimated the maximal Lyapunov exponent this way for the
GOY model with standard parameters
$N=19, \nu = 10^{-6}, k_0 = 2^{-4}, f = (1+i)*0.005$. 
The value of the time increment was varied from $\delta t$ = 0.001 all the
way up to $\delta t$ =10, in time units set by the choice of parameters
and we find that the results are independent of this value, as expected.
Fig. 1 shows the obtained results, with a variation
of  $\epsilon$ 
in the interval [0,2]. Note the transition to chaos at a
critical value $\epsilon_c = 0.38704$ \cite{com}. Biferale et al \cite{bif}
found a slightly larger value, $\epsilon = 0.398$, for the transition 
to the attractor at the same parameter values; we do not know 
the origin of this discrepancy.
Magnifying around the transition point we observe a finite
but small jump of $\lambda$ from 0 up to 0.015. This makes the transition
first order but we speculate that this jump might disappear in the
$N \to \infty$ limit. After that the Lyapunov exponent grows more
or less linearly with $\epsilon$ and reaches a maximum value
$\lambda \sim 0.54$ at $\epsilon = 0.92$. Then it drops sharply down
to zero around $\epsilon \simeq 1$ after which it raises
and drops again. The $\epsilon = 1$ is special because at this point the
last term in the GOY model, which couples to the two previous shells, is zero.
This means that for the last three shells, the equations reduce to:
\begin{eqnarray}
(\frac{d}{ dt}+\nu k_{N-2}^2 ) \ u_{N-2} \ & = &  i \,k_{N-2} 
( u^*_{N-1}u^*_N -\frac12 u^*_{N-3}u^*_{N-1}) \nonumber \\
(\frac{d}{ dt}+\nu k_{N-1}^2 ) \ u_{N-1} \ & = & 
- i \,k_{N-2} (u^*_{N-2} u^*_N )\\
(\frac{d}{ dt}+\nu k_N^2 ) \ u_N \ & = & 0 \nonumber 
\end{eqnarray}
It is seen that $u_N$ decays exponentially to zero, and therefore 
we may neglect the
right-hand-side of the second equation, which then leads to an exponential
decay of $u_{N-1}$. Just the same argument can be applied to the
behavior of $u_{N-2}$ up to $u_4$ where the forcing is applied. 
We thus expect that the solution essentially converges towards the trivial
fixed point $u_n=0$ \cite{Fridolin} and {\it not} to 
the ``Kolmogorov fixed point''. 
This is in accordance with our numerical results, where we see a fast 
exponential convergence of the modes toward $u_n=0$. 
Fig. 2 shows the variation of the amplitudes $| u_n |$ as a function
of time. Besides the nice exponential decay, it appears that
each mode is successively ``triggered'' for the decay. 
This is seen in Fig.2: the n'th shell is triggered when the
absolute value of the amplitude of the $n+1$'th shell is a
approximately $5 \cdot 10^4$ and the amplitude of the $n+2$'th shell has
practically vanished. This means that the equation for the n'th shell becomes
$\frac{d}{dt} u_n = -i\,k_{n-1}(u^*_{n-1} u^*_{n+1}) - \nu k_n^2\ u_n$, and 
just when the $n+1$'th amplitude reaches $ \sim \ 5 \cdot 10^4$ the 
right-hand-side will be dominated by the viscosity which 
then forces the n'th amplitude towards zero. 

Because the system is forced on the 4'th shell, the amplitude
of this shell is prevented from going to zero. In fact, when
the forcing is included the ``trivial fixed point'' is 
$ u_n^* = (0,0,0,{f \over {\nu k_4^2}},0,0,..... )$.  Fig. 3 shows 
the variation of the real part of the maximal eigenvalue
of the Jacobian as function of $\epsilon$ evaluated in this
fixed point. We observe that the fixed point is repelling
for the standard parameter value $\epsilon=0.5$ but becomes
attractive for $\epsilon=1.0$. This is also the place
where the dynamics becomes non-chaotic as indicated by a
convergence (although very slow) of the Lyapunov exponent towards zero.

In order to understand the above results further we try to formulate the
simplest possible model with the same ``symmetries'' as the GOY 
model, a shell map.
This map is constructed from the same principles as the GOY model
with the difference that the time interval between each update is
a full time unit. We then obtain the following map
in the real ``velocity'' variable $v_n (i)$ of shell $n$ 
at integer time $i$:
\begin{equation}
\label{vn}
 v_n (i+1) \ =  v_n (i) ~+~
 k_n (A_n \,   v_{n+1} (i) v_{n+2} (i) \, + \, \frac{B_n}{2} 
v_{n-1} (i) v_{n+1} (i) \, + \,
 \frac{C_n }{4} \,   v_{n-1} (i) v_{n-2} (i) )~-~ \nu_1 k_n^2 \ v_n (i)
+ \ f_1 \delta_{n,1}
\end{equation}
with boundary conditions $B_1=B_N=C_1=C_2=A_{N-1}=A_N=0$.
Conservation of the energy $E=\sum_n v_n(i)^2$ tells us that,
for $\nu_1 = f_1 =0$,
the differential $d E=2 \sum_n v_n(i) d v_n(i) \simeq
2 \sum_n v_n(i) \cdot (v_n(i+1) -v_n(i)) = 0$ leading to
a similar relations between the coefficients as for the
GOY model:
$ A_n=1,\ B_n=-\epsilon_1,\ C_n=-(1-\epsilon_1)$ \cite{note}. 
Surprisingly enough, iterations of this model is stable even
for very few shells.

To simplify, we start out with a shell map with only $N = 5$ shells
and the following parameters: $\nu = 10^{-4}, k_0 = 2^{-1}, f = 0.00005$.
It is immediate to study the eigenvalues of the Jacobian at
the fixed point $u_n(i+1) = u_n(i)$. 
We locate the fixed point in
the stable regime by iterating the map and refining the solution with 
Newton's method. This fixed point is a ``Kolmogorov fixed point''
with a scaling close to $v_n \sim k_n^{-1/3}$, apart oscillations
\cite{bif}. By varying the parameter $\epsilon_1$ in small steps
and then using Newton's method again one can obtain the fixed
point where the map becomes unstable. 
We calculate the eigenvalues of the Jacobian at the fixed point and 
observe a Hopf-bifurcation (a pair of two complex conjugate eigenvalues 
escapes from the unit disc in the complex plane) at 
$\epsilon_1=0.705081885$.
Here a limit cycle appears. From numerical iteration we observe that the 
limit cycle has a period of 217.822 (iterations)
at $\epsilon_1 = 0.70509$. At this 
point, the phase of the eigenvalues equals 0.0288458 which is 
in agreement 
with the numerically found period since $2\cdot \pi / 0.0288458 = 217.820$.
Fig. 4a shows the limit cycle for 
the parameter value $\epsilon_1 =0.73$ (note the 
somewhat unusual ``return plot'' where the $n+1$'th shell
amplitude is plotted against the $n$'th shell amplitude. This gives
nicer graphs that a usual return plot). As $\epsilon_1$ is increased
further a series of standard period doubling bifurcations occur,
Fig. 4b shows period four at $\epsilon_1 = 0.764$. Finally, the 
the motion ends up on a strange
attractor as shown in Fig. 4c for $\epsilon_1 = 0.77$.

To estimate the maximal Lyapunov exponent for the shell map
a similar technique as for the GOY model is applied except that now
we are dealing with a discrete map.
The tangent vector $\tilde{\bf z} (i)$ is propagated according to
\begin{equation}
{\tilde z}_k (i+1) = \sum_{j=1}^{N} {\tilde A}_{kj} \cdot 
{\tilde z}_j
\end{equation}
where ${\tilde A}_{kj}$ is the Jacobian of the map (\ref{vn})
\begin{equation}
{\tilde A}_{kj} ~=~ \partial v_k(i+1)  /  \partial v_j(i)
\end{equation}
Again, practically we initiate by a unit tangent vector in
a random direction ${\bf z}(0)$ and follow the expansion of
the eigenvector. The only difference to the GOY model is that the time 
increment now is unity so Eq. (\ref{eq5}) applies with $\delta t =1$.

Also at low values of $\epsilon_1$ there is a transition and
Fig. 5 show the variation of the maximal Lyapunov exponent as
a function of $\epsilon_1$.
For $\epsilon_1 \simeq 0$
the model starts out to be chaotic but then the Lyapunov exponent
drops and becomes negative around $\epsilon_1 \simeq 0.15$. The
exponent approaches zero at $\epsilon_1 \simeq 0.705$. We identify this
transition with the appearance of the limit cycle: the distance
between two initially close trajectories on the limit cycle will not
converge nor diverge. Then
the transition to chaos described above
occurs at $\epsilon_{1_c} = 0.766$. 
After that $\lambda$
grows sharply and drops back to zero at $\epsilon_1 \simeq 1$.
It is interesting that there seem to be windows where $\lambda$ goes to
zero, which are like the windows found in the logistic map
(see for instance \cite{Ott}). 
At $\epsilon_1$ = 1  we find again that the system converges towards the
fixed point close to zero 
$(\frac{f}{\nu k_1^2},0,0,0,0)$ (and not the Kolmogorov fixed point). 
The eigenvalues of the
Jacobian are in this case just $1-\nu k_i^2$ ($i=1,\dots 5$) and thus
the fixed point is stable.

It is possible to apply the shell map also for more shells and
one can go up to $N =11$ without divergencies. Fig. 6 shows
the variation of the maximal Lyapunov exponent as a function 
of $\epsilon_1$. Here the transition occurs at $\epsilon_{1_c} =
0.657$
For this number of shells one can identify an inertial range
and below the transition, the structure functions follow the
Kolmogorov predictions, although with the standard oscillations.
Above $\epsilon_{1_c}$ the motion is intermittent and we again identify 
windows of stable evolution as for the $N=5$ case.

J.K. is grateful to the Niels Bohr Institute for warm hospitality.

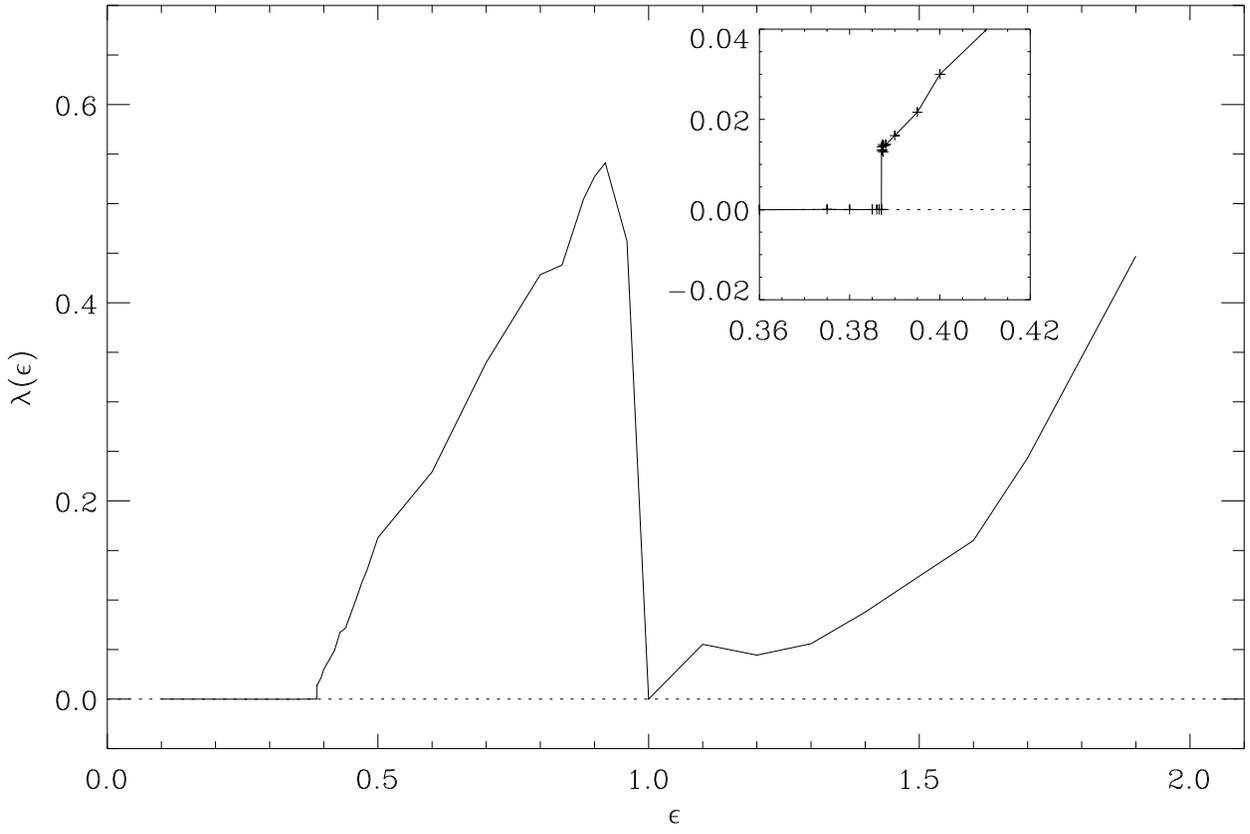
\begin{figure}
\caption[x]{The variation of the maximal Lyapunov exponent as
a function of $\epsilon$ for the GOY model equations (\ref{un}) with
the parameters $N=19, \nu = 10^{-6}, k_0 = 2^{-4}, f = (1+i)*0.005$.
Note the transition to chaos at $\epsilon_c = .38704$ and the finite, but 
small jump at this point. The Lyapunov exponent increases nearly linearly
but drops back almost to zero at $\epsilon =1$ as explained in the text.
The ``bump'' at $\epsilon =1.1$ is due to
the fact that the value of $\lambda$ has not converged.
\label{fig:1}
}
\end{figure}

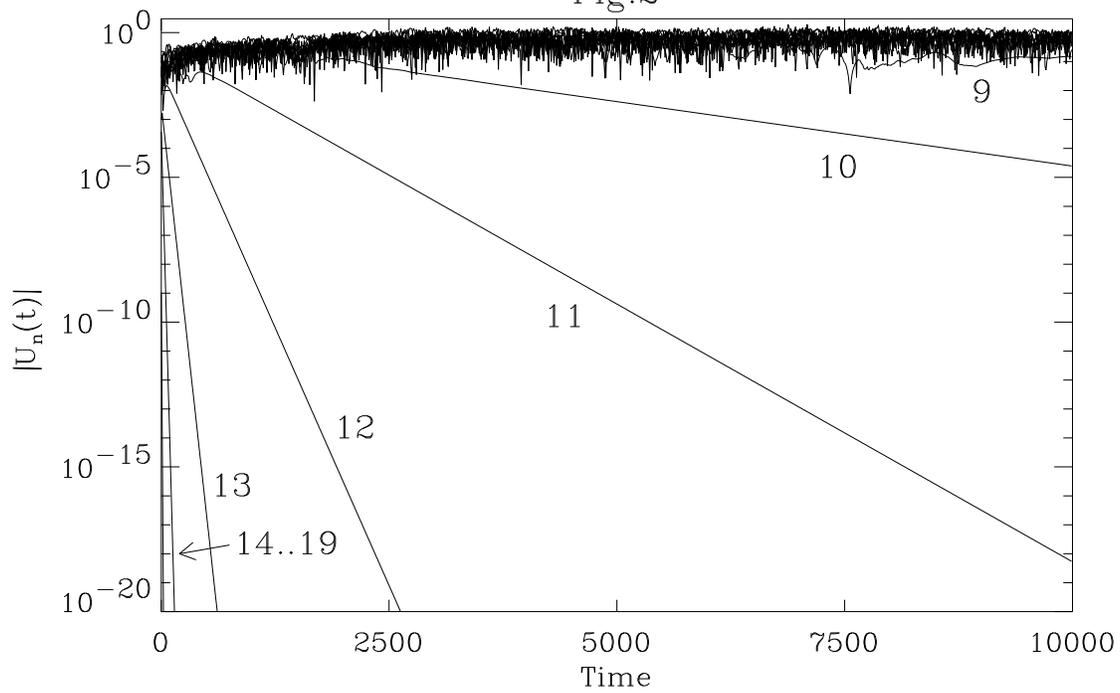
\begin{figure}
\caption[x]{A logarithmic plot of the long-term dynamics of 
the absolute value of the shell amplitudes  
$| u_n |$ for $\epsilon=1$. The labelling refers to the shell number
and one notes the drastic decay the of the higher shell amplitudes.
\label{fig:2}
}
\end{figure}

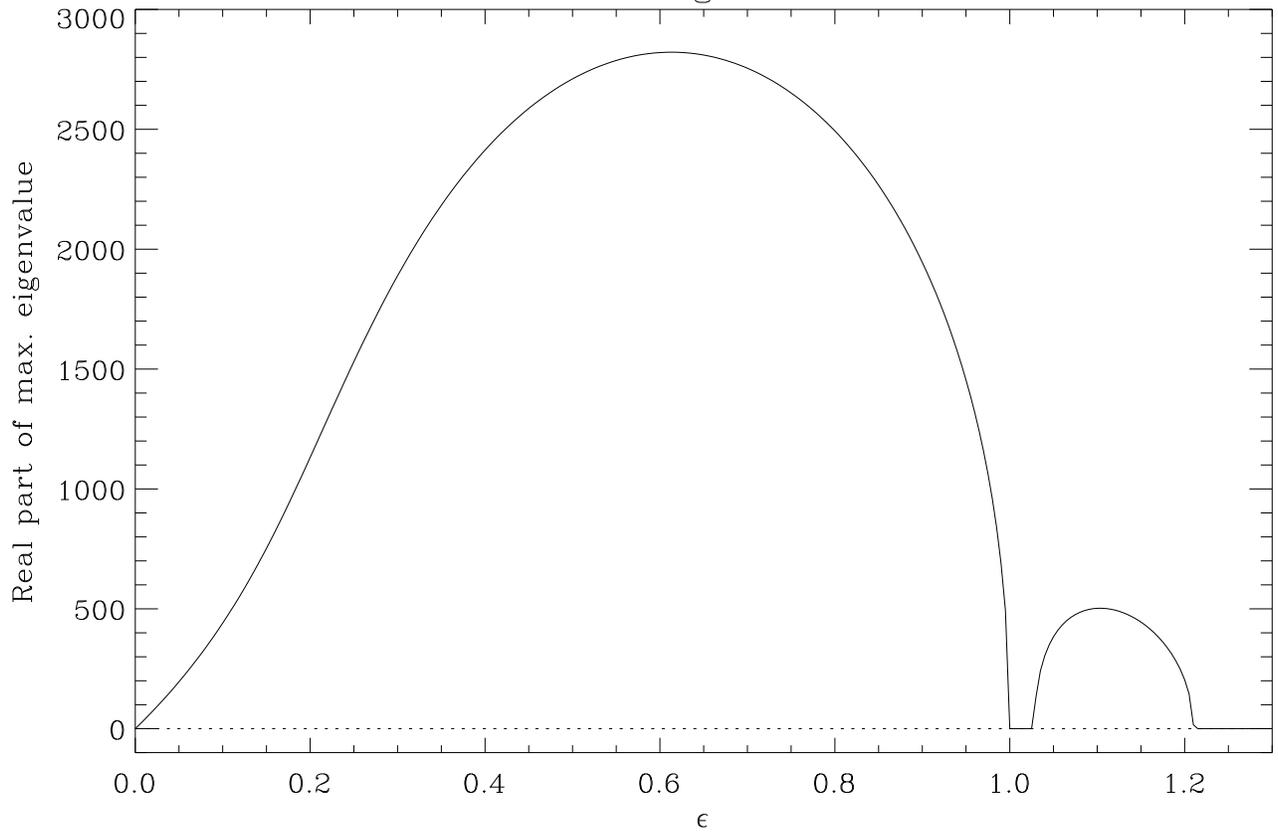
\begin{figure}
\caption[x]{The value of the real part of the maximal eigenvalue
of the Jacobian evaluated in the ``trivial fixed point'' 
$ u_n^* = (0,0,0,{f \over {\nu k_4^2}},0,0,..... )$, 
as a function of $\epsilon$.
Note, that the fixed point becomes stable around $\epsilon =1$.
\label{fig:3}
}
\end{figure}

\begin{figure}
\caption[x]{``Return plots'' $(v_n(i),v_{n+1}(i))$
of the shell map Eq. (\ref{vn}) with
the parameters $N= 5, \nu = 10^{-4}, k_0 = 2^{-1}, f = 0.00005$.
In a) $\epsilon_1 = 0.73$ and we observe limit cycle generated by
the Hopf bifurcation; in b) the second period doubling of the limit
cycle is shown at $\epsilon_1 = 0.764$ and c) is the strange attractor
for $\epsilon_1 = 0.77$. 
\label{fig:4}
}
\end{figure}
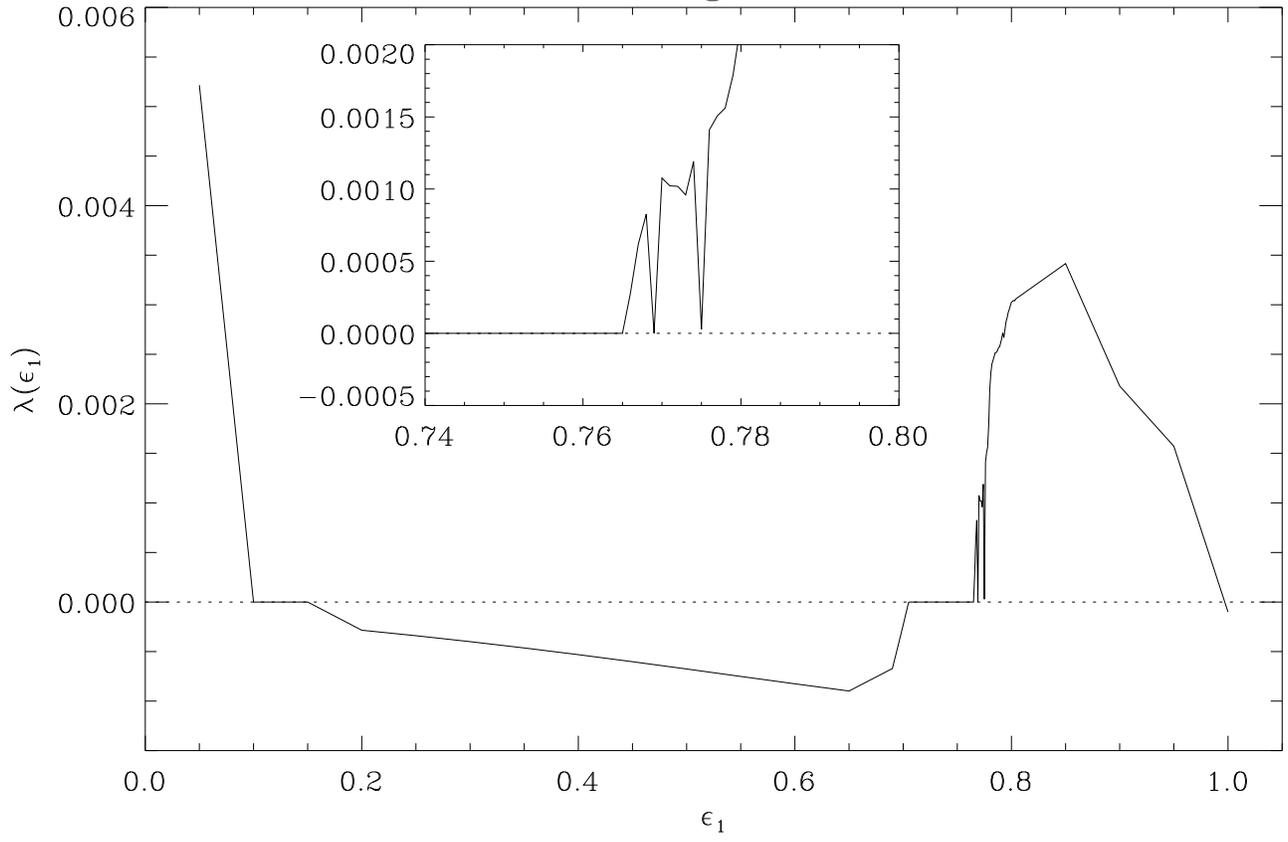
\begin{figure}
\caption[x]{The maximal Lyapunov exponent versus $\epsilon_1$
for the shell map
in Eq. (\ref{vn}) for the parameters 
$N= 5, \nu = 10^{-4}, k_0 = 2^{-1}, f = 0.00005$. Note the two transitions
and that the model becomes regular again for $\epsilon_1 = 1$.
\label{fig:5}
}
\end{figure}

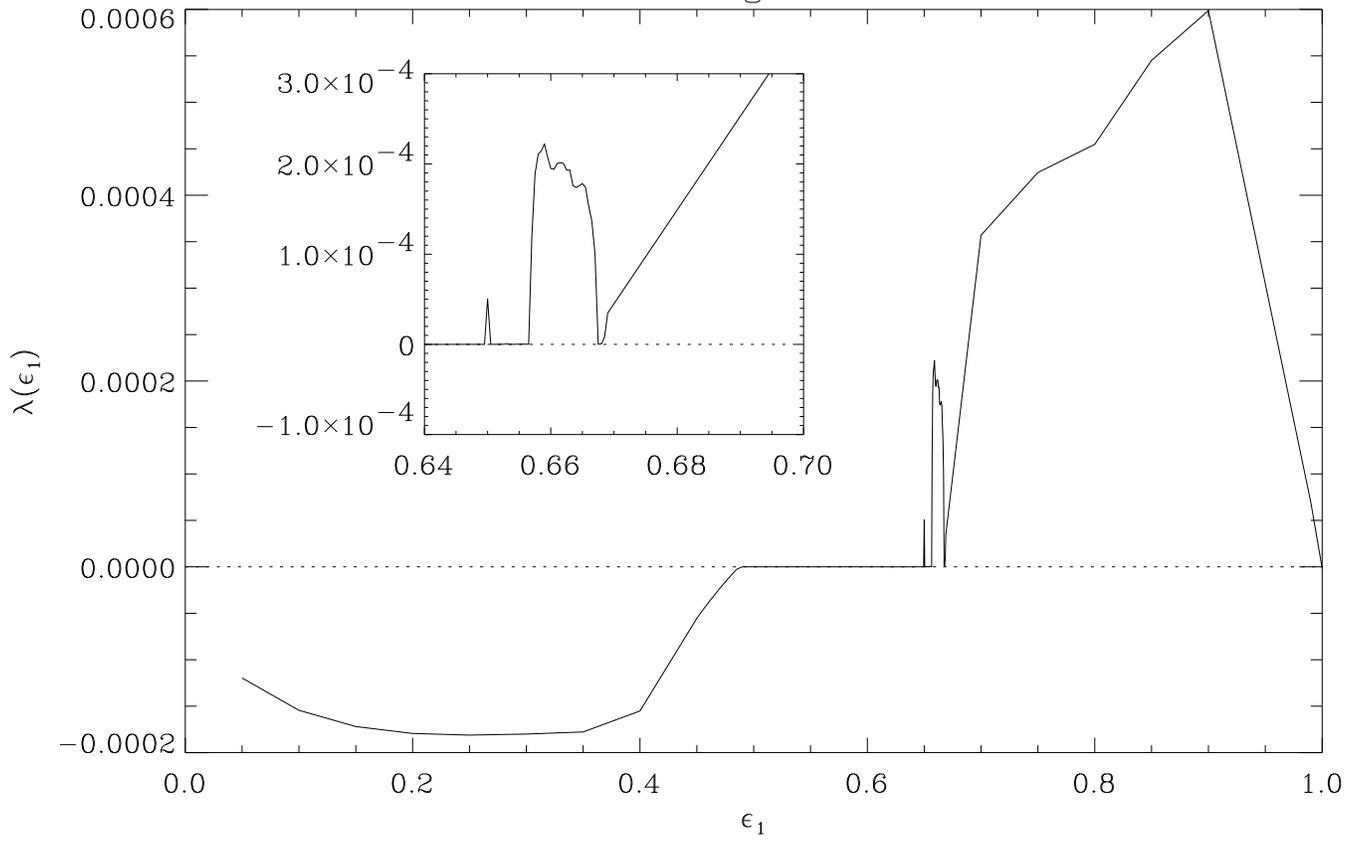
\begin{figure}
\caption[x]{The maximal Lyapunov exponent versus $\epsilon_1$
for the shell map
in Eq. (\ref{vn}) for the parameters
$N=11 , \nu = k_0 = 2^{-4}, f = 5 \cdot 10^{-6}$. Here, there
is only one transition at $\epsilon_{1_c} \sim 0.66$
\label{fig:6}
}
\end{figure}


\begin{thebibliography}{10}
\bibitem{com}
The critical value of $\epsilon$ for the transition depends on
the values of the parameters and also on the valueof $N$ although we believe
it converges for large $N$.
\bibitem{note}
Strictly speaking, energy conservation means:
$0 ~=~ \sum_n (v_n(i+1)^2 - v_n(i)^2) ~=~ \sum_n
(v_n(i+1) + v_n(i)) \cdot (v_n(i+1) - v_n(i))$. We see in (\ref{vn}) that in the
limit $\nu_1 = f_1 = 0$ this is fulfilled for the stationary Kolmogorov
solution $v_n \sim k_n^{(-1/3)}$.
\end{thebibliography}
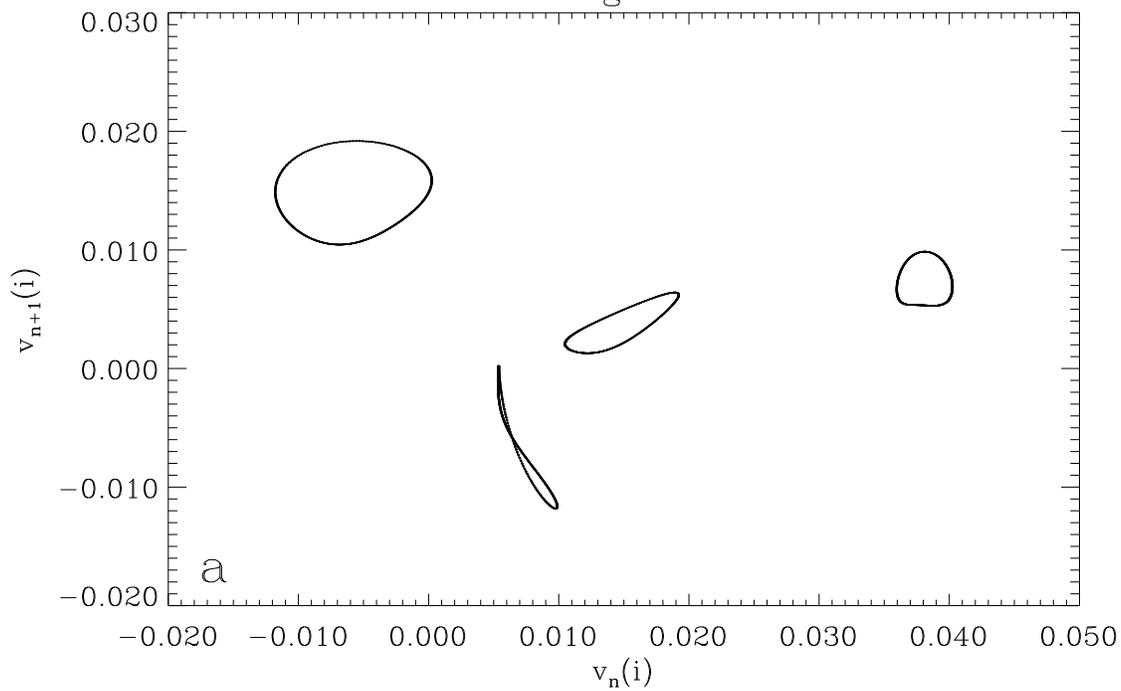
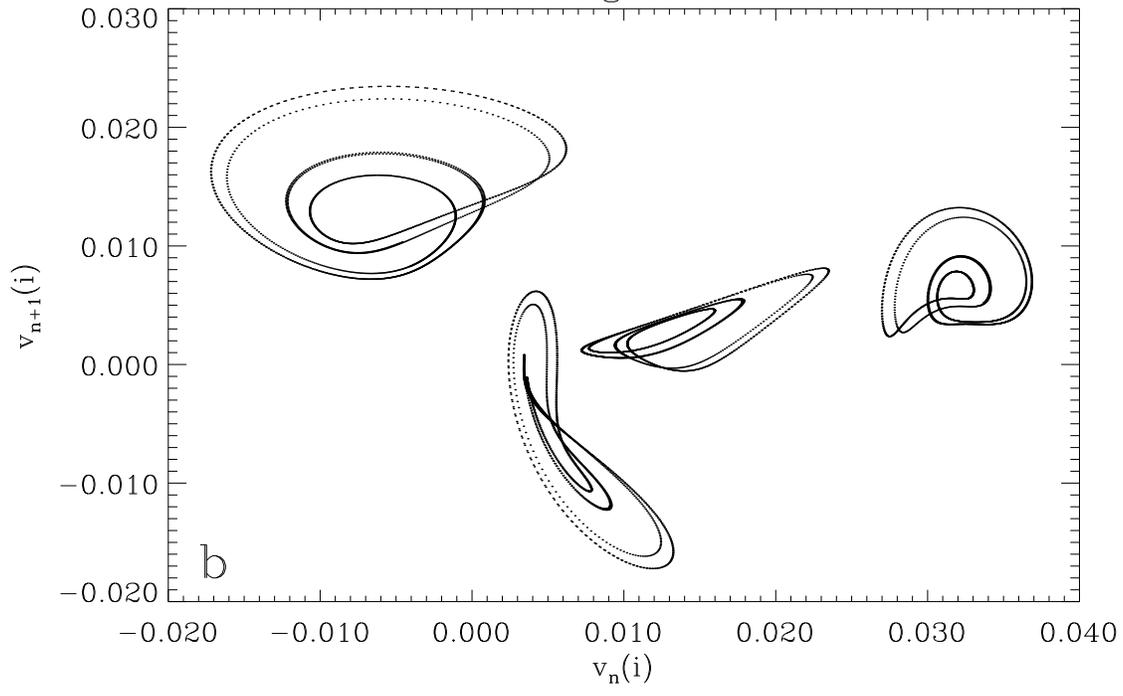
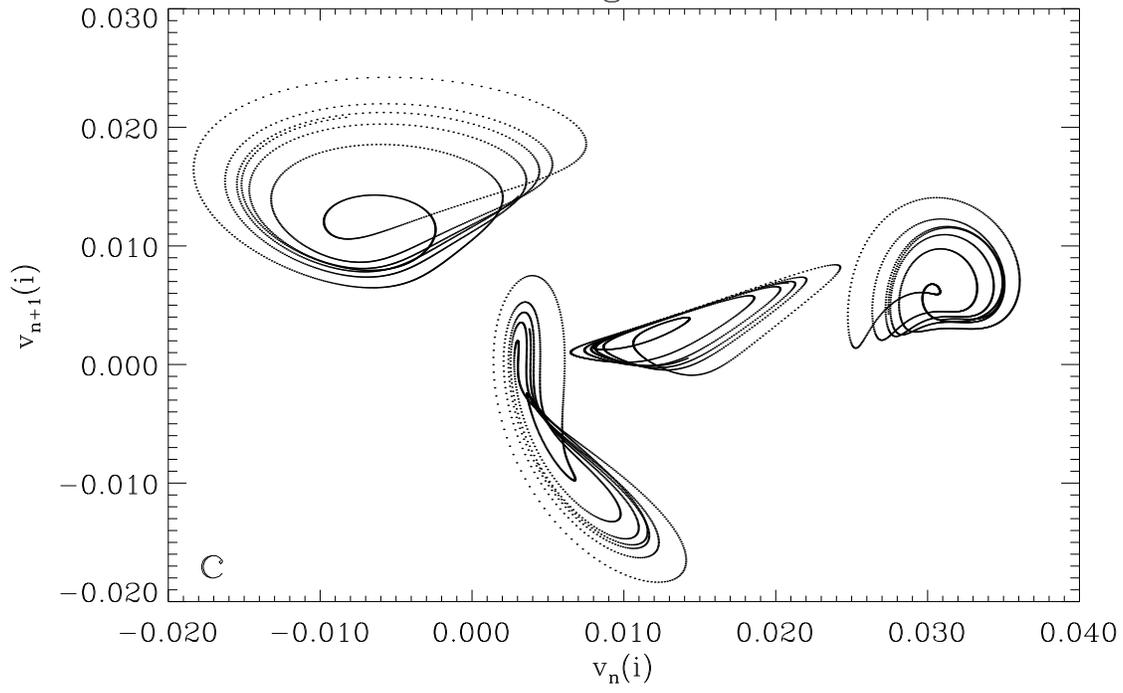
\end{document}